\documentclass[
prl, twocolumn,
superscriptaddress,
nofootinbib,
 amsmath,amssymb,
 aps,
floatfix,
]{revtex4-1}

\usepackage{graphicx}
\usepackage{textcomp} 
\usepackage{gensymb} 
\usepackage{dcolumn}
\usepackage{bm}
\usepackage{amsmath, amssymb}
\usepackage{dsfont}
\usepackage{float}
\usepackage{afterpage}
\usepackage{xcolor}
\usepackage[normalem]{ulem}

\begin{document}



\title{Hiding Ignorance Using High Dimensions}

\author{M. J. Kewming} 
\email{michael.kewming@gmail.com}
\affiliation{Centre for Engineered Quantum Systems, School of Mathematics and Physics, University of Queensland, QLD 4072 Australia}
\author{S. Shrapnel}
\affiliation{Centre for Engineered Quantum Systems, School of Mathematics and Physics, University of Queensland, QLD 4072 Australia}
\author{A. G. White}
\affiliation{Centre for Engineered Quantum Systems, School of Mathematics and Physics, University of Queensland, QLD 4072 Australia}
\author{J. Romero}
\email{jacq.romero@gmail.com}
\affiliation{Centre for Engineered Quantum Systems, School of Mathematics and Physics, University of Queensland, QLD 4072 Australia}

\date{\today}
\begin{abstract}
The absence of information---entirely or partly---is called ignorance. Naturally, one might ask if some ignorance of a whole system will imply some ignorance of its parts. Our classical intuition tells us yes, however quantum theory tells us no: it is possible to encode information in a quantum system so that despite some ignorance of the whole, it is impossible to identify the unknown part \cite{vidick_does_2011}. Experimentally verifying this counter-intuitive fact requires controlling and measuring quantum systems of high dimension $(d {>} 9)$. We provide this experimental evidence using the transverse spatial modes of light, a powerful resource for testing high dimensional quantum phenomenon. 
\end{abstract}

\maketitle

Entropic inequalities have found use in a wide variety of practical settings in physics, including non-locality \cite{oppenheim_uncertainty_2010}, information causality \cite{barnum_entropy_2010}, cryptography \cite{damgaard_tight_2006,wehner_cryptography_2008,tomamichel_tight_2012,furrer_continuous-variable_2018}, and quantum memories \cite{berta_uncertainty_2010}. Vidick and Wehner (VW) quantify a dimensional-dependent entropic inequality which holds for all measurement non-contextual hidden variable models (NC-HV) \cite{vidick_does_2011}. The inequality states that, in a NC-HV model of a composite system made of two parts, the ignorance in the whole can be clearly identified in at least one of its parts. Quantum mechanics is contextual \cite{spekkens_contextuality_2005} and hence violates the VW-inequality highlighting the counter-intuitive properties of quantum information stored in systems comprising of parts. It has been an open question whether it is possible to verify this inequality experimentally. Here we do so finding that in quantum systems ignorance of the whole does not imply ignorance of the parts.

Violation of the VW-inequality is challenging as it requires a large degree of control over the preparation and measurement of high-dimensional qudits. Intriguingly, it depends on dimension, only being violated for systems of dimension $d{>}9$. In this work, we use the transverse spatial profile of photons, a widely-used and readily accessible qudit basis \cite{forbes_creation_2016}. This degree of freedom has been used successfully to demonstrate the Einstein-Podolsky-Rosen paradox \cite{leach_quantum_2010}, to violate the Bell inequality in higher dimensions \cite{dada_experimental_2011}, for intra-city quantum cryptography \cite{sit_high-dimensional_2017}, and for free-space communication \cite{krenn_orbital_2017}. 

\begin{figure}[!b]
    \centering
    \includegraphics[width=\columnwidth]{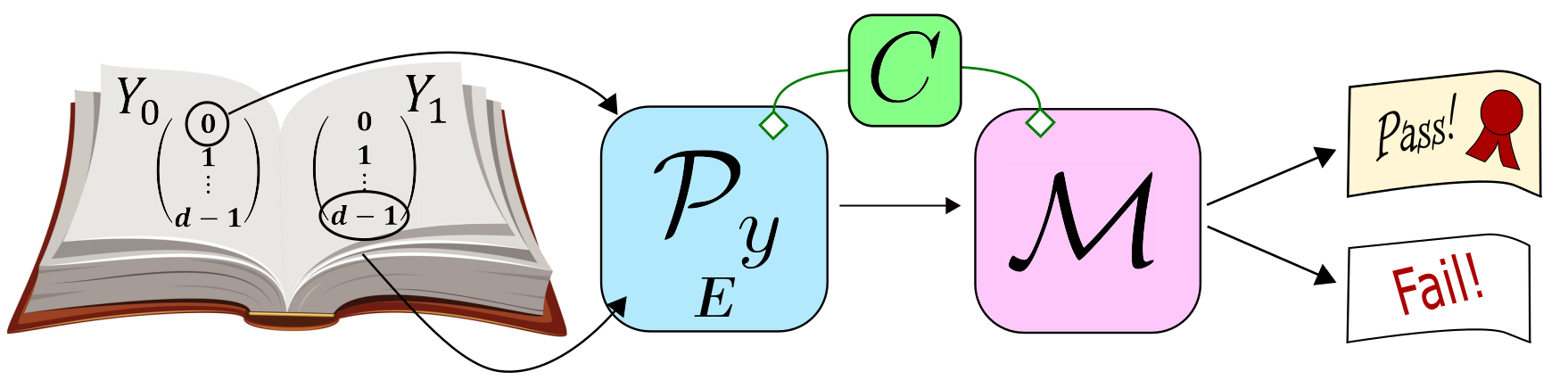}
    \caption{A textbook contains two chapters with information $Y_0$ and $Y_1$. An ignorant student---who has not know any content---will be asked to answer a question from either chapter. Luckily they are allowed to bring a single answer---1 dit---corresponding to either $y_{0} {\in} Y_{0}$ or $y_{1} {\in} Y_{1}$ via study notes $\mathcal{P}_{y}$ prepared by their knowledgeable friend (who knows all $Y$). These notes are stored physically in a register $E$. In the test, the student can be asked to answer a single question $\mathcal{M}$ about the whole $y=y_{0}y_{1}$, or the parts $y_{0}$ or $y_{1}$. The teacher has access to a system $C$ that is classically correlated with the student's notes (green line). The teacher's goal is to ask questions $\mathcal{M}$ such that she can uncover which chapter the student is ignorant of. (See supplementary materials for more details).
    }
    \label{fig:conception}
\end{figure}
 
The conceptual significance of the VW-inequality can be appreciated via the following analogy: consider a student who has not studied for an exam about the contents of a two-chapter textbook. Each chapter contains $d$ possible questions with an equivalent number of answers. The student is required to know the answers $y_{0}$ and $y_{1}$ to two possible questions, uniformly drawn from either chapter i.e $y_{0} \in Y_{0}$ and $y_{1} \in Y_{1}$. The combination of answers from both chapters is a single dit string $y {=} y_{0}y_{1}$, uniformly drawn from the Cartesian product of the random variables $Y = Y_{0} \times Y_{1}$. Before the test the student is permitted to store \emph{only one dit} of information in their study notes $\mathcal{P}_{y}$. We can model $\mathcal{P}_{y}$ as a physical preparation, where we assume the possibility of some hidden variables that determine the state of the preparation. That is, $\mathcal{P}_{y}$ is assumed to be sampled from a distribution over all hidden variables $\lambda$. We can imagine a knowledgeable friend who knows all the content $Y$ and what questions could be asked on the exam, and who prepares the study notes for the student. Consequently, regardless of the question being asked the student will only have knowledge of the single dit of information their friend encoded in $E$ but not the particular $\lambda$ associated with this instantiation of $\mathcal{P}_{y}$. In the test, the student can be asked to answer a single question $\mathcal{M}$ about the whole $y=y_{0}y_{1}$, or the parts $y_{0}$ or $y_{1}$, giving the conditional probability of successfully answering $y$ as $P(y{|}\mathcal{M}, \mathcal{P}_{y})$. 

The expected probability of the student guessing the whole book $Y$---or either part $Y_{0}$ or $Y_{1}$---is computed as a weighted sum
\begin{equation}
p_{\text{guess}}(Y|E) {=} \underset{\lbrace \mathcal{M} \rbrace}{\max}\sum_{y} P_{Y}(y)P(y|\mathcal{M}, \mathcal{P}_{y}) \,, 
\end{equation}
where $P_{Y}(y){=}1/d^{2}$ (recall, $y_{0}$ and $y_{1}$ are selected at random). The maximisation over $\mathcal{M}$ signifies the highest success probability of the student guessing $Y$ correctly. The \emph{ignorance} of $Y$ is defined as the minimum entropy achievable, conditioned on any possible knowledge gained via the physical encoding $E$: $H_{\infty}(Y|E) {=} {-} \log p_{\mathrm{guess}}(Y|E)$ \cite{konig_operational_2009}.  In base 2, we can interpret the conditional min-entropy as the maximum length of a bitstring that is uniquely determined by $Y$ and independent of $E$ \cite{konig_sampling_2011}.


This ignorance of the whole $Y$ is bounded by the dimension of the encoded system $d$ because knowing the whole requires knowing both dits. Given the encoding is only a single dit then one dit of information cannot be encoded, hence $H_{\infty}(Y|E) {=} \log d$ \cite{vidick_does_2011}. In fact, this bound also holds in the quantum case where the student's notes are replaced by a single \emph{qudit} \cite{kraus_complementary_1987,maassen_generalized_1988}.

Consider how the ignorance of the parts relates to this ignorance of the whole. If the student prepares their notes such that they always encode the answers from $Y_{1}$; $E {=} Y_{1}$. As a result they know $Y_{1}$ with certainty $P(Y_{1}|E{=}Y_{1}){=}1$ but must guess the $Y_{0}$ randomly $P(Y_{0}|E{=}Y_{1}){=}1/d$. For $d{=}2$, the probability of guessing the parts correctly is $p_{\mathrm{guess}}(Y_{C}|E){=}0.75$, where $C$ is a classical random variable $c{\in}\lbrace 0,1\rbrace$ pointing to the part that must be answered. Assuming the teacher has access to the system $C$---which is classically correlated with the student's notes $E$ such that $P(c{=}0{|}E) {=} 1$---then they will always ask a question from $Y_{0}$, forcing the student to guess randomly. Thus, system $C$ points to the source of ignorance. 

This is the intuition behind the VW-inequality: For a given encoding $E$ of any dimension $d$, there always exists a random variable $C$ that can point to the source of ignorance 
\begin{equation}
\label{eq:min_entropy}
    H_{\infty}(Y_{C}|E,C) \geq \frac{H_{\infty}(Y|E)}{2} - 1\,,
\end{equation}
where $H_{\infty}(Y|E)$ is the \emph{ignorance of the whole} and $H_{\infty}(Y_{C}|E,C)$ is the \emph{ignorance of the parts}. One can think of this inequality as a constraint on how ignorance of the whole $H_{\infty}(Y|E)$ can be split between the two parts $Y_{C}$, derived from the min-entropy splitting lemma \cite{damgaard_tight_2006,wullschleger_oblivious-transfer_2007}. 
Since we are encoding two random dits in a single qudit ($H_{\infty}(Y|E) {=} \log d$) then the inequality can be rewritten $p_{\mathrm{guess}}(Y_{C}|E,C) \leq 2/\sqrt{d}$. The $2$ in the numerator arises from the $-1$ in Eq.\,(\ref{eq:min_entropy}) accounting for the single bit pointing variable $C$ which can be classically correlated with $E$. Here $p_{\mathrm{guess}}(Y_{C}|E,C)$ determines the amount of possible randomness extracted---or conversely, the information gain---from $Y_{C}$ conditioned on $E$ and $C$. This can be intuitively understood in the following way: If the inequality is satisfied, then the observed randomness in $Y_{C}$ can be explained by a NC-HV model. That is, the randomness introduced by the preparation strategy $E$ can be explained by a distribution $\mathcal{P}_{y}$ over hidden variables $\lambda$ which the pointing variable $C$ effectively uncovers. If the inequality is violated for all $C$, then the randomness extracted from $Y_{C}$ is lower than expected and cannot be explained by a NC-HV model. This implies distributions of $Y_{0}$ and $Y_{1}$ \emph{cannot} be independently separated from one another once they have been encoded and extracted from $E$.


Interestingly, if there is an \emph{additional} source of ``helpful'' noise which leaks $m$ extra bits of information from student to teacher, then the inequality is modified via the chain rule whereby the R.H.S acquires an additional $-m$ \cite{vidick_does_2011}. This leads to $p_{\mathrm{guess}}(Y_{C}|E,C) \leq 2^{1+m}/\sqrt{d}$, making the inequality exponentially more difficult to violate; there is more randomness that must be discounted. Here we experimentally assume this additional noise is zero $m{=}0$.


\begin{figure*}
    \centering
    \includegraphics[width=2\columnwidth]{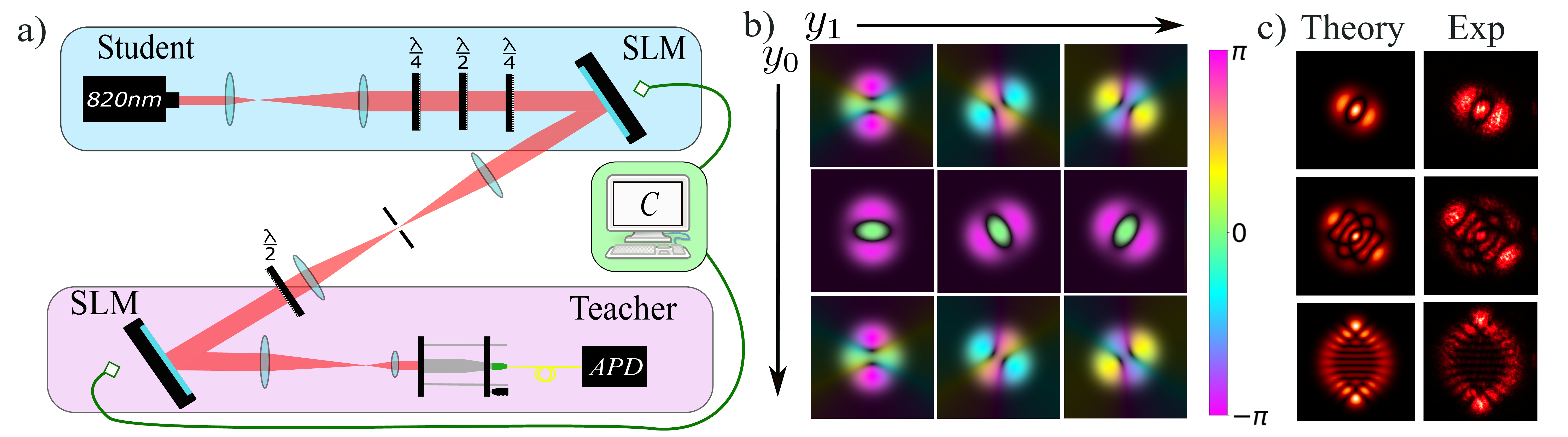}
    \caption{a) In the experiment, the student's notes consists of a single qudit $\rho_y^E$ encoded in the spatial profile of a single photon. The photon is prepared using a phase mask displayed on a spatial light modulator (SLM). The photon then propagates to the second SLM through a $4f$-lens configuration. The teacher sets the phase mask of the second SLM, corresponding to a question $M_{y_{c}}$, where $c$ indicates the chapter from which the teacher's question is derived. The photon is then coupled to a single-mode fibre connected to single-photon counting module, where a photon detected signifies a correct answer. The distribution $C$ is correlated with the student's notes by a classical computer. b) Each image shows the spatial amplitude overlayed with the phase profile for each unique answer for the case of $d{=}3$. c) We can produce the encoded modes experimentally with high fidelity. Here we have plotted experimental intensity plots of several randomly chosen encodings for $d=3,7$, and $11$ compared with ideal case.}
    \label{fig:modes}
\end{figure*}

If we replace the encoded dit with a qudit $\rho_{y}^{E} {=} \vert \Psi_{y} \rangle \langle \Psi_{y} \vert$. The questions become positive-operator valued measurements (POVM) $\sum_{y} M_{y} {=} \mathbb{I}$. Without loss of generality, the random variable $C$ can be represented as a classical mixed state $\sigma_{y}^{C} {=} q_{y} \vert 0 \rangle \langle 0 \vert + (1- q_{y})\vert 1 \rangle \langle 1 \vert$, e.g a biased coin, with $q_{y}$ a measure of the bias. Similarly, the probability of successfully guessing $y$ becomes $P(y|\mathcal{M}, \mathcal{P}_{y}) {=} \mathrm{tr}\left(\rho_{y}^{E} M_{y}\right)$.

The student encodes the answers $y_{0}$ and $y_{1}$ into a single qudit---a single quanta---using two mutually unbiased bases, the generalised Pauli operators $X_{d}$ and $Z_{d}$
\begin{equation}
\label{eq:encoding}
    \vert \Psi_{y} \rangle {=} \frac{X_{d}^{y_{0}}Z_{d}^{y_{1}}\left(\mathbb{I} {+} F\right)}{\sqrt{2\left(1{+}1/\sqrt{d}\right)}}\vert{0}\rangle\,,
\end{equation}
where $F$ is the quantum Fourier transform. In this encoding,  measurements in the computational basis (eigenbasis of $Z_{d}$) reveals $y_0$ and measurements in the Fourier basis (eigenbasis of $X_{d}$) reveals $y_1$. In this encoding, the answers from $Y_{0}$ are encoded in the Fourier eigenbasis of $X_{d}$ and answers from $Y_{1}$ are encoded in the computational eigenbasis of $Z_{d}$. These states are referred to as the maximally-certain states \cite{oppenheim_uncertainty_2010} and used in many quantum security protocols including BB84 \cite{nielsen_quantum_2011}, random access codes \cite{crepeau_quantum_1994, tavakoli_quantum_2015}, and oblivious transfer \cite{wullschleger_oblivious-transfer_2007,chailloux_optimal_2016}.


Using this type of encoding, both $Y_{0}$ and $Y_{1}$ can be answered with equal probability ${p_{\text{guess}}(Y_{c}|E,C) {=} (1 + 1/\sqrt{d})/2}$. For $d{=}2$ we also have $p_{\text{guess}}(Y_{c}|E,C){\approx}0.853$ gaining an advantage over the classical case of 0.75, i.e. there is less randomness if we use the encoding in Eq.\,(\ref{eq:encoding}).    Furthermore, guessing the parts is now independent of the system $\sigma_{y}^{C}$, that is $C$ can no longer point to the source of ignorance. Consider increasing the dimension: At $d{=}8$, the RHS of the inequality implies we can communicate $0.5$ bits of randomness whereas the LHS implies $0.56$ bits---the inequality is satisfied. Now consider the case of $d{=}10$, where the inequality is violated. From the RHS, we expect the parts to have at least $0.66$ bits of randomness. Using the Eq.\,(\ref{eq:encoding}) we measure $0.60$ bits of randomness for \emph{both} parts (LHS). Assuming $E$ is drawn from a distribution $\mathcal{P}_{y}$ over hidden variables $\lambda$, the randomness is lower in the parts than is predicted. The pointing variable $C$ cannot uncover this distribution and furthermore, cannot explain the reduction in randomness of the parts. We must conclude the quantum encoding $E$ cannot be explained by a NC-HV; hence ignorance of the whole does not imply ignorance of the parts \cite{vidick_does_2011}.

Violation of (\ref{eq:min_entropy}) requires a versatile experimental platform capable of supporting high-dimensional Hilbert spaces with a large degree of control over preparations and measurements. We use the transverse spatial modes of light, here described in terms of the Laguerre-Gauss bases.  Each mode in this basis ($LG_{l,p}$) is fully characterised by two numbers $\{l,p\}$. The encoded state  $\vert \Psi_{y} \rangle$ is represented as a weighted superposition of the modes, $\vert \Psi_{y} \rangle {=} \sum_{i} a_{i} \vert \psi \rangle _{l_{i},p_{i}} $, where $a_{i}$ is a complex number.

Large qudit systems can be produced and measured using phase masks displayed on liquid-crystal spatial light modulators (SLMs) \cite{molina-terriza_twisted_2007, langford_measuring_2004, forbes_creation_2016}. Our experimental apparatus is depicted in Fig. \ref{fig:modes}(a). The first SLM is used to prepare a \emph{d-rail} qudit---where each rail is an orthogonal LG mode---from an incoming photon initially in the mode $LG_{0,0}$. The photon then propagates through a $4f$-lens configuration and is re-imaged on the second SLM where a measurement mask $\vert \Phi_{y} \rangle$ is displayed. Re-imaging the prepared state $\vert \Psi_{y} \rangle$ onto the second SLM allows us to measure the overlap $\langle \Phi_y \vert \Psi_{y}\rangle$. The resulting beam is coupled to a single-mode fibre connected to a single-photon detector. The count rate in the detector is then proportional to the overlap $\langle \Phi_y \vert \Psi_{y}\rangle$. The guessing probability---and subsequently the min-entropy---are obtained using
\begin{equation}
    \mathrm{tr}\left(\rho_{y}^{E}\,M_{y}\right) = \frac{\vert \langle \Phi_{y}\vert \Psi_{y} \rangle \vert^{2}}{\vert \langle \Psi_{y}\vert \Psi_{y} \rangle \vert^{2}} = \frac{O_{y}}{N_{y}} \,.
\end{equation}
Here $N_{y}$ is the photon counts when we prepare and measure the $\vert \Psi_{y} \rangle$, providing a normalisation for our measurement counts $O_{y}$. If a photon is detected, this corresponds to the student answering the question correctly. The pointer variable $C$ is realised by correlating the choice of measurement and encoding phase masks using a classical system i.e a computer. 

In our experiment, the spatial profile of the single photon carries the information about the dit string $y$. In (\ref{eq:encoding}), $y_{0}$ is encoded in the computational basis corresponding to the amplitude, $|a_i|$, of each mode $\vert \psi \rangle _{l_{i},p_{i}}$. On the other hand, $y_{1}$ is encoded in the relative phases between each mode, $\arg({a_i})$. The physical characteristics of this encoding can be seen in Fig\,.\ref{fig:modes}(b) for the nine possible configurations of $y$ when $d=3$.

We use a highly attenuated $809\,$nm CW beam in the $LG_{0,0}$ mode with a mean photon number of $\vert \alpha \vert^{2} {\sim} 0.01$ as our input. On average we detect ${\sim}10^{6}$ photons/s using a Perkin Elmer SPCM-AQR single photon counting module with an average dark count of ${\sim}150\,$photons/s. We use two Meadowlark 1920 x 1152 analog spatial light modulators to display the preparation and measurement phase masks. We use the amplitude modulation technique described in Ref. \cite{bolduc_exact_2013} to calculate the phase masks. For a uniform intensity on the illuminated region of the first SLM, the input beam is made much larger than the encoding phase mask. In Fig.\,\ref{fig:modes}(c), we show ideal and experimentally derived intensities for several random encodings of $y$---up to $d{=}11$. The amplitude modulation introduces a mode-dependent reflection efficiency that we account for in the normalisation of our statistics \cite{bolduc_exact_2013}.  

\begin{figure}
    \centering
    \includegraphics[width = \columnwidth]{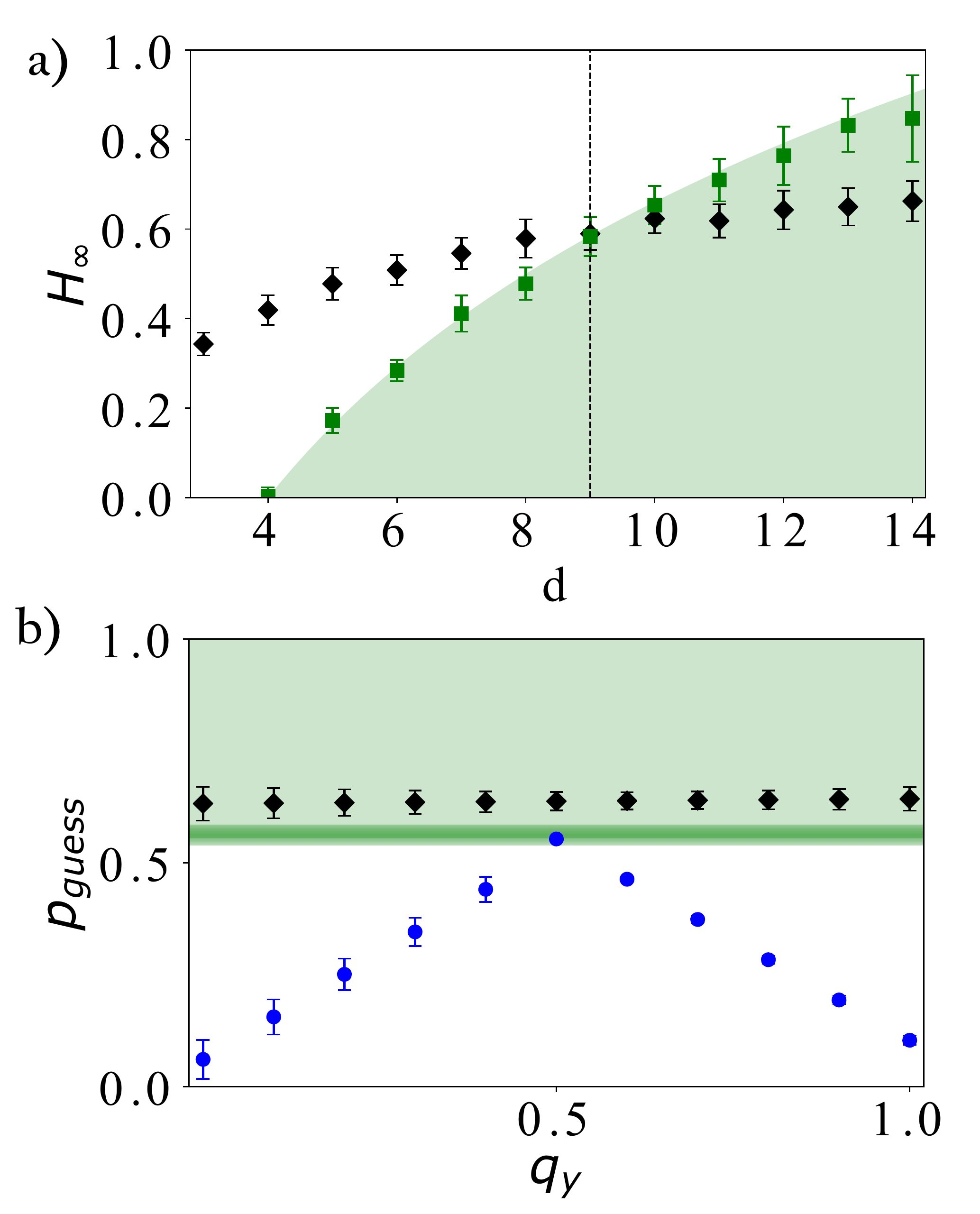}
    \caption{a) Experimental results of the student's ignorance---plotted as min-entropy $H_{\infty}$---for increasing dimension $d$. The black diamonds are experimental values for the student's ignorance of the \emph{parts} $H_{\infty}(Y_{c}|E,C)$ using the quantum encoding $\rho_{y}^{E}$. The green squares show the student's ignorance of the \emph{whole} $H_{\infty}(Y|E)$. At $d>9$ we observe that the student's ignorance of the parts is lower than the ignorance of the whole. b) We choose $d{=}13$ to check if the teacher's system $\sigma_{y}^{C}$ can uncover which part the student is ignorant of. We plot the guessing probability rather than min-entropy. Recall $\sigma_{y}^{C}$ is a classical mixed state like a biased coin, where $q_{y}$ indicates the bias. If $q_{y}{=}1$ then the teacher always asks the student for the answer to $Y_{0}$, and vice versa. We observe that using $\rho_{y}^{E}$, the student's guessing probability of the parts (black diamonds) is constant and is always greater than the whole (green band) for \emph{any} $\sigma_{y}^{C}$. This confirms the violation of the VW-inequality.
    For the optimal classical case (blue) we can always find a $\sigma_{y}^{C}$ that points to the unknown part. Here we plot the guessing probability for $E{=}Y_{1}$ for $q_{y}<0.5$ and $E{=}Y_{0}$ for for $q_{y}>0.5$. Error bars correspond to one standard error.}
    \label{fig:data}
\end{figure}

Initially, we measured the entropy of both sides of (\ref{eq:min_entropy}) as a function of the dimension $d$, without accounting for the possibility of a classically correlated system $\sigma_{y}^{C}$. To collect a full set of statistics, we repeated the experiment for every combination of $y$ from $2{\leq} d {\leq} 14$ and measured in both the $X_{d}$ and $Z_{d}$ bases. We collect $2d$ measurements for each of the $d^{2}$ encodings yielding a total of $2d^3$ unique measurements for each $d$. From this complete data set, we can reconstruct the min-entropy of the whole $Y$ and both of its parts $Y_{0}$ and $Y_{1}$. Our results agree closely with the theoretical predictions, shown in Fig.\ref{fig:data}(a). We demonstrate that using the encoded states in (\ref{eq:encoding}) for $d{>}9$ the RHS of the inequality---ignorance of the whole---becomes greater than the LHS---ignorance of the parts. 

Satisfying Eq.\,(\ref{eq:min_entropy}) requires that the source of ignorance can be uncovered for \emph{some} binary distributions $\sigma_{y}^{C}$. Experimentally it is infeasible to test all possible distributions over hidden variables $\lambda$. We note this is true for all experimental tests of measurement non-contextuality \cite{mazurek_experimental_2016}. Consequently, we establish that it is impossible to satisfy Eq.\,(\ref{eq:min_entropy}) by choosing $d{=}13$---where the difference between the parts and the whole is greatest for our measurements. We randomly choose measurements according to the mixed distribution $\sigma_{y}^{C}$. The VW-inequality is satisfied for any data point that is in the white area of Fig. \ref{fig:data}(b). By varying the bias $q_{y}$ of the binary distribution $\sigma_{y}^{C}$, we show that it is impossible to satisfy the VW-inequality using the encoded state given by (\ref{eq:encoding}) (black diamonds). Thus we show that ignorance of the whole does not imply ignorance of the parts. 
We also experimentally verify that classical systems will always satisfy (\ref{eq:min_entropy}). We test the optimal classical strategy, where $E{=}Y_{0}$ (right) or $E{=}Y_{1}$ (left). In both these examples, the results show a strong dependence on the choice of system $\sigma_{y}^{C}$ as shown in Fig.\ref{fig:data}(b) and satisfy (\ref{eq:min_entropy}).

Inequalities in physical theories have played a significant role in distinguishing between the quantum and classical world. Here we experimentally demonstrate that \emph{ignorance of the whole does not imply ignorance of the parts}, thus highlighting yet another distinction between the quantum and classical world. Our result stems from the fact that observed probabilities in quantum mechanics are contextual---any model satisfying Eq.\,(\ref{eq:min_entropy}) must be non-contextual. 

Our work provides a flexible architecture where we can control quantum systems of high dimensionality. We focused on a game-like, adversarial scenario in this work, but our system is applicable to general communication scenarios where a quantum system is being transmitted from sender to receiver. Communication by transmitting high-dimensional quantum system has been related to entanglement-assisted classical communication where dimension is a parameter \cite{tavakoli_dimensional_2017}. Recent results \cite{tavakoli_dimensional_2017, martinez_high-dimensional_2018} have shown the onset of an advantage above a certain dimension---the intuition for this boundary is still lacking. In contrast, our boundary can be understood by considering the interplay of dimensionality and noise. If their are additional ``helpful'' bits of noise---a non-zero leakage of $m$ bits between transmitter and receiver---the dimension for violating Eq.\,(\ref{eq:min_entropy}) scales exponentially as $d=(2^{m+2}-1)^{2}$. For $m{=}0$,  using the quantum encoding Eq.\,(\ref{eq:encoding}) with $d>9$, leads to a randomness in the parts that is less than that predicted by a NC-HV model.  Investigating the interplay of dimension with other elements such as quantum correlations or noise could be a fruitful direction for understanding dimension-dependent boundaries in other quantum communication scenarios \cite{tavakoli_dimensional_2017, martinez_high-dimensional_2018, coles_entropic_2017}.  

\emph{Acknowledgments}
This work is supported by the Australian Research Council Centre of Excellence for Engineered Quantum Systems (EQUS, CE170100009). JR is supported by an ARC Discovery Early Career Research Award (DE160100409) and a L'Oreal-UNESCO For Women In Science Fellowship Award. AGW is supported by a University of Queensland Vice-Chancellor Research and Teaching Fellowship. JR would like to thank Thomas Vidick and Stephanie Wehner for initial discussions.
We acknowledge Gerard Milburn and Marco Tomamichel for fruitful discussions and suggestions.  

\bibliography{Ignorance}

\newpage
\section*{Supplementary material}

\subsection*{The teacher/student example}
In this article we have described the conceptual significance of the theory using a detailed narrative of a student sitting a series of tests. Here we present a table relating all the symbols to each object in the narrative to aid the reader.
\begin{table*}
\begin{tabular}{||c|c|c|c||}
\hline
    Narrative & Classical & Quantum & Dimension \\
\hline
    Book & $Y$ & $Y$ & $d^{2}$\\
    Answers to each chapter & $Y_{0}$, $Y_{1}$ & $Y_{0}$, $Y_{1}$ & d\\
    Possible answers to test & $y$ &$y$ & $d^{2}$\\
    Individual answers &$y_{0}$, $y_{1}$ &$y_{0}$, $y_{1}$  & $d$\\
    Study notes & $\mathcal{P}_{y}$ &$\rho_{y}^{E}$  & $d$\\
    Question & $\mathcal{M}$ &$M_{y}$  & $d$ \\
    Outcome of test & $P(y\vert \mathcal{M}, \mathrm{P_{y}})$ &$\mathrm{tr}(\rho_{y}^{E} M_{y})$ & -\\ 
    Teacher's correlated system & $C$ &$\sigma_{y}^{C}$ & $2$\\
 \hline 
\end{tabular}
\caption{This table shows the object in the narrative and the corresponding mathematical object in both the classical and quantum case. We also include the dimension of the system. }
\label{Tab:table}
\end{table*}
Common language can sometimes unintentionally obscure the more nuanced aspects of the problem. Here is a much more brief and direct overview of the protocol.
\begin{itemize}
    \item A random dit string of length 2 is selected uniformly at random $y=y_{0}y_{1}$. 
    \item It must be encoded into a single qudit $\rho_{y}^{E}$ in the register $E$. This process erases some information, therefore it is impossible to know the whole $y$. 
    \item The qudit is then measured using the positive operator valued measurement POVM $M_{y}$. The measurement operator $M_{y}$ is maximised to reveal either the parts $y_{0}$ or $y_{1}$ or the whole $y$ depending on the test. 
    \item The probability of successfully guessing $y$ is then given by the trace operator $\mathrm{tr}(\rho_{y}^{E} M_{y})$.
    \item To obtain a measurement of the whole $Y$, we sum over all combinations of $y$ multiplying each probability of the outcome by the probability that $y$ was selected. Hence $p_\mathrm{guess}(Y|E) = \underset{\lbrace M_{y} \rbrace}{\max} \sum_{y} P_{Y}(y)\mathrm{tr}(\rho_{y}^{E} M_{y})$ where $P_{Y}(y)=1/d^{2}$. Similarly, we can measure the parts in the same way e.g $p_{\mathrm{guess}}(Y_{0}|E) = \underset{\lbrace M_{y_{0}} \rbrace}{\max} \sum_{y} P_{Y_{0}}(y_{0})tr(\rho_{y}^{E} M_{y_{0}})$ where $P_{Y}(y)=1/d$.
    \item The min-entropy $H_{\infty}(Y|E) = - \log p_{\mathrm{guess}}(Y|E)$ can then be computed by taking the log of each guessing probability.
\end{itemize}

\subsection*{The encoding}
The encoding used to violate the inequality is an equal superposition of two mutually unbiased bases (MUB)
\begin{equation}
    \vert \Psi_{y} \rangle {=} \frac{1}{\sqrt{2\left(1+1/\sqrt{d}\right)}}X_{d}^{y_{0}}Z_{d}^{y_{1}}\left(\mathbb{I} + F\right)\vert{0}\rangle
\end{equation}
where $F$ is the quantum Fourier transform and $X_{d}$ and $Z_{d}$ are the generalised Pauli operators that are the generators of the Heisenberg-Weyl group. They are defined as
\begin{equation}
    X_{d} \vert y_{0} \rangle = \vert y_{0}+1\,\mathrm{mod} \,d \rangle \quad \mathrm{and} \quad Z_{d} \vert y_{0} \rangle = \omega^{y_{0}} \vert y_{0} \rangle\,,
\end{equation}
where $\omega = \exp \left(2\pi i /d\right)$. The Pauli operators form a canonical conjugate pair and are related by $Z_{d} = F^{\dag} X_{d} F$. In quantum mechanics the notion of canonically conjugate quantities is central irrespective of Hilbert space dimension. If the state of a system is such that one canonical variable takes a definite value, then the conjugate must be maximally uncertain. In the original proof of the VW-inequality, the authors assumed $d$ was prime to complete the proof \cite{vidick_does_2011}. This assumption is required due to the unanswered questions relating to MUBs for non-prime dimensions \cite{durt_mutually_2010}. Currently, it is not known how many MUBs exist for composite Hilbert dimensions, but is well known for prime powers and the continuous limit.

This type of encoding is needed to create a quantum superposition, of the $X_{d}$ and $Z_{d}$ eigenstates. The inequality holds in all non-contextual-hidden variable models where $E$ is a classical distribution. Hence, violation of the VW-inequality can only be achieved if $E$ is not in a deterministic combination of $X_{d}$ and $Z_{d}$. 

\subsection*{$d$-rail qudits}
Here we present an overview of the $d$-rail qudits analysis. Let $d$ represent the number of available modes, then the total Hilbert space is the tensor product of Fock space spanned by the states 
\begin{equation}
    \vert n_{1}, n_{2}, ..., n_{d} \rangle \equiv \vert n_{1} \rangle \otimes \vert n_{2} \rangle \otimes ... \otimes \vert n_{d} \rangle\,.
\end{equation}
We will further assume that we are working in a subspace in which every state is an eigenstate of the total photon number operator $\hat{N} \vert \psi \rangle = N \vert \psi \rangle$ where $\hat{N} = \sum_{j=1}^{d} a_{j}^{\dag}a_{j}$ where $N$ is a positive integer and $a_{j}^{\dag},a_{j}$ are the creation and annihilation operators which satisfy $[a_{i},a_{j}^{\dag}] = \delta_{ij}$. A single photon $N=1$ has a two dimensional Hilbert space and two modes spanned by the Fock states $\lbrace | 1,0 \rangle, | 0,1\rangle\rbrace$, often called a dual-rail qubit \cite{kok_linear_2007}. Here, we consider the case of a single photon $N=1$ photon with $d$ modes, hence a Hilbert space of dimension $d$ will be spanned by the Fock states of one photon in each mode.

A reliable single-photon source is required to produce the Fock states described above. We will show that a weak single photon coherent state is sufficient for our purposes. In the case of a single mode, a coherent state is defined as 
\begin{equation}
    \vert \alpha \rangle = e^{-\vert \alpha \vert^{2}/2} \sum_{n=0}^{\infty} \frac{\alpha^{n}}{\sqrt{n!}} \vert n \rangle\,,
\end{equation}
where $\alpha$ is an arbitrary complex number. Coherent states do not have a fixed photon number but the average is bounded and equal to $\vert \alpha \vert^{2} = \langle n \rangle$. We generate $d$ modes each in a coherent state 
\begin{equation}
\label{eq:state}
\vert \Psi \rangle = \vert \alpha_{1} \rangle \otimes \vert \alpha_{2} \rangle \otimes...\otimes \vert \alpha_{d} \rangle = \vert \alpha_{1}, \alpha_{2},..., \alpha_{d} \rangle\,.
\end{equation}
We can see that this state will be contaminated by undesired multi-photon Fock states. Furthermore linear optics can only transform coherent product states into coherent products states: no entanglement is possible. 

It is possible to get around this restriction by introducing an imaginary measurement device that makes a total photon measurement on $d$ modes without absorbing any photons or mode mixing. We can now ask: what is the conditional state if such a measurement is made on a $d$-fold product of coherent states, conditioned on the result of the measurement being $N$? We can write this conditional state as 
\begin{equation}
\label{eq:conditional}
    \vert \Psi : N \rangle = \mathcal{N} \hat{\Pi}_{N} \vert \alpha_{1}, \alpha_{2},..., \alpha_{d} \rangle\,,
\end{equation}
where $\hat{\Pi}_{N}$ is the projection operator onto the subspace of total photon number $N$ and $\mathcal{N} = p_{N}^{-1/2}$.
Because all our modes are generated at the SLM from the same input beam, we will make the assumption that the coefficients of each mode is equal up to multiplicative factor $\alpha_{i} = \alpha \beta_{i}$ where $\vert \beta_{i} \vert < 1$ and satisfies $\sum_{j=1}^{d}|\beta_{i}|^{2} = 1$. We can interpret $|\beta_{i}|^{2}$ as the probability amplitude associated with each of the $d$ modes.
This allows us to define the probability of having $N$ photons in the experiment as 
\begin{equation}
    p_{N} = \mathrm{tr}\left( \vert \Psi : N \rangle \langle \Psi: N \vert\right) = \frac{\vert \alpha\vert ^{2N}}{N!}e^{-\vert \alpha \vert^{2}}  \,,
\end{equation}
which is equivalent to the single mode case as we would expect. 

We are interested in the single photon case $N=1$ where $\hat{\Pi}_{1} = \vert 1 \rangle\langle 1\vert$. The probability of measuring a single photon state is $p_{1} = e^{-\vert \alpha \vert^{2}}\vert \alpha \vert^{2}$. In our experiment, vacuum states are not counted and do not contribute to the statistics. We measure a mean photon number $\vert \alpha \vert^{2} \sim 0.01$ which means $p_{1}\approx0.01$ and $p_{2}\approx5\times10^{-5}$. For every 60,000 we would expect 3 photons i.e $~17$ instances per $10^{6}$ counts which is negligible. Therefore, will only consider the possibility that we have up to two photons in the experiment.

These two photon events will create two types of errors in our counting statistics. The first error occurs when two photons were in the experiment but only one was counted i.e photon loss. This is simulated as using a beam splitter model with transmissivity $\eta$ standing in for the role of non-unit quantum efficiency. Such a filter is a linear optical device that transforms the input annihilation operator $a_{k}\rightarrow \sqrt{\eta}a + \sqrt{1-\eta} b_{k}$ where $b_{k}$ is an auxiliary mode that is initially in the vacuum state. The second error occurs when two photons are both detected and show up as a single event. We do not use number resolving detectors so we cannot distinguish photon number and must account for it.

Our input beam is a highly attenuated coherent state $\vert \alpha \vert^{2} \ll 1$ where we will truncate to only include two photon modes at most. In this limit, our product state can be approximated by 
\begin{equation}
    \vert \Psi \rangle \approx e^{-\vert \alpha \vert^{2}/2}\left(1+ \alpha\sum_{i=1}^{d} \beta_{i}a_{i}^{\dagger} + \frac{\alpha^{2}}{2}\left[\sum_{i=1}^{d}\beta_{i} a_{i}^{\dagger}\right]^{2}\right)\bigotimes_{i}^{n}\vert 0\rangle_{i}\,.
\end{equation}
Making the beam splitter transformations $\vert \Psi \rangle \rightarrow \vert \Psi' \rangle$ we obtain 
\begin{widetext}
\begin{equation}
        \vert \Psi'\rangle= e^{-\vert \alpha \vert^{2}/2}\left(1 + \alpha\sum_{i=1}^{d} \beta_{i} \left(\sqrt{\eta}a_{i}^{\dagger} + \sqrt{1-\eta}b_{i}^{\dagger}\right) + \frac{\alpha^{2}}{2}\left[\sum_{j=1}^{d}\, \left(\sqrt{\eta}a_{i}^{\dagger} + \sqrt{1-\eta}b_{i}^{\dagger}\right)\right]^{2} \right)
    \bigotimes_{i}^{d}\vert 0 \rangle_{a,i}\vert 0 \rangle_{b,i}\,,
\end{equation}
\end{widetext}
Because we cannot resolve the photon number, we will not be able to distinguish between one and two photon events. As a result, we will have the mixed distribution over conditional states defined in (\ref{eq:conditional})
\begin{equation}
    \rho_{N} = p\vert \psi':1 \rangle\langle \psi':1\vert + (1-p)\vert \psi':2 \rangle\langle \psi':2\vert\,,
\end{equation}
where we ignore the vacuum state because it never enters our counting statistics. 
Here $p=\left(1+ \vert \alpha \vert^{2} /2\right)$ and can be interpreted as probability we measured a single photon, conditioned on the event that a detection was made.
We also cannot measure the lost photons in the $b$ mode and must partially trace out this system of $\rho_{N}$. We again condition on the events where there was a detection in and the $a$ mode---dropping all independent $b$ modes---leading us to a final mixed state,
\begin{equation}
    \rho = p\vert \Psi \rangle\langle \Psi \vert + (1-p)\left[\eta \vert \Psi\rangle \langle \Psi \vert_{2a} + (1-\eta) \vert \Psi\rangle \langle \Psi \vert_{ab}\right]\,.
\end{equation}
Here $\vert \Psi \rangle$ is our desired single photon state;  $\vert \Psi \rangle_{2a}$ corresponds to the case when 2 were detected as a single event; $\vert \Psi \rangle_{ab}$ is the case where one photon was correctly detected but the other was lost. These last two states will corrupt our statistics.

\begin{figure}[b]
    \centering
    \includegraphics[width=\columnwidth]{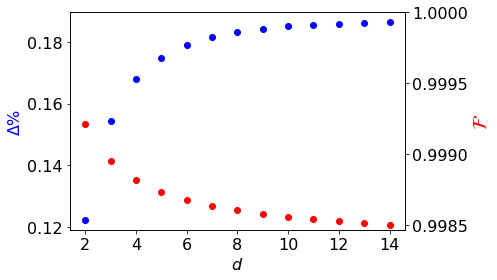}
    \caption{The difference in guessing probability $\Delta\%$ (blue) and the fidelity $\mathcal{F}$ (Red) of the desired state with the mixed state as a function of dimension. The guessing probability is only minutely corrupted by the presence of the contamination states in the single photon limit for a coherent state with $\vert \alpha \vert^{2}=0.01$ and increases with dimension but begins to taper off. A similar behaviour is seen in the fidelity, begining high but then decreasing with dimension.  }
    \label{fig:measure}
\end{figure}

We can now compute the effect that non-unit quantum efficiency $\eta$ has on our results using the encoding $\vert \Psi_{y} \rangle$ shown for $\vert \alpha \vert^{2} = 0.01$ and $\eta = 0.6$---the approximate quantum efficiency of our APD---in Fig \ref{fig:measure}. Here we plot the effect that the contamination states have on the absolute difference in the guessing probability using the ideal vs the mixed states $\Delta\% = \vert P_{\mathrm{guess}} - P_{\mathrm{guess}}'\vert$. The contamination states introduce an extremely small error---on the order of $\sim ~-0.2 \%$--- to our measurement statistics. This can easily be understood by computing the fidelity $\mathcal{F}$ of the mixed state we observe in the lab to our desired state which is also show in Fig \ref{fig:measure}.

\end{document}